\documentstyle[11pt]{article}
\newcommand{\be}{\begin{equation}}
\newcommand{\ee}{\end{equation}}
\newcommand{\bea}{\begin{eqnarray}}
\newcommand{\eea}{\end{eqnarray}}
\newcommand{\del}{\partial}

\newcommand{\e}{\epsilon}

\newcommand{\vs}[1]{\vspace{#1 mm}}
\newcommand{\hs}[1]{\hspace{#1 mm}}

\begin{document}

\baselineskip=.60cm

\thispagestyle{empty}

\rightline{hep-th 0203239}

\vs{20}

\centerline{\large\bf World-Volume Description of M2-branes Ending on 
an M5-brane}
\centerline{\large\bf and Holography} 
\vs{15}
\centerline{N.S. Deger\footnote{e-mail: deger@gursey.gov.tr} and A. 
Kaya\footnote{e-mail: kaya@gursey.gov.tr}}
\vs{5}
\centerline{Feza Gursey Institute, Emek Mah. No:68, 81220, Cengelkoy,
Istanbul, TURKEY}
\vs{25}

\begin{abstract}
We consider world-volume description of M2-branes ending on an M5-brane.
The system can be described either as a solitonic solution of the M5-brane
field equations or in terms of an effective string propagating in
6-dimensions. We show that the zeroth order scalar scattering amplitudes
behave similarly in both pictures. The soliton solution appears to have a
horizon-like throat region. Due to the underlying geometric structure of
the M5-brane theory, modes propagating near the horizon are subject to a
large red-shift. This allows one to define a decoupling limit and implies
a holographic duality between two theories which do not contain dynamical
gravity.
\end{abstract}

\newpage

\setcounter{page}{1}

\section{Introduction}

\hs{5}There is now considerable evidence that quantum gravitational
theories in $d$-dimensions on $AdS$ spaces are dual to conformal gauge
theories defined in $d-1$ dimensions \cite{maldacena, gubser, witten}.
This is a
concrete realization of
the holographic principle \cite{hooft, susskind} which states that in
quantum gravity, physics within a volume can be described in terms of a
theory without gravity on the boundary of that volume. The holographic
principle is usually associated with the presence of dynamical gravity in
the bulk, and indeed the original motivation was to explain the area
dependence of
the black hole entropy. In this letter, we will argue that this principle
can be generalized to world-volume theories of branes, which indicates
that
holography is mainly related to {\it diffeomorphism} invariance acting as
a gauge symmetry. Although, world-volume theories do not contain dynamical
gravity, there is still the notion of the {\it induced}
geometry and gravity. A possible holographic duality of this type, between
the 6-dimensional open membrane (OM) theory on $AdS_3\times S^3$ and
${\cal N}=(4,4)$ superconformal field theory in 2-dimensions, has
 been proposed in \cite{perr} which motivated this paper.

\
\

We consider the (2,0) theory in 6-dimensions \cite{witm51, witm52} which is
associated with the M5-brane dynamics. The theory is conformally invariant;
there is no (dimensionless) coupling constant and no length scale
\cite{seib1}. There are 16 real supersymmetries and the supermultiplet
consists of a self-dual 2-form potential, 5 scalars and chiral 
spinors. The moduli
space labeling the vacua of the theory is given by ${\cal R}^{5K}/W$,
where
$K$ is the number of coincident M5-branes and $W$ is a discrete group
\cite{seib1}. The fundamental degrees of freedom are presumably related to
self-dual, tensionless string excitations in 6-dimensions, and thus it is
hard to give a field theoretic definition of the theory. On the other hand,
it can be obtained from M-theory by taking a decoupling limit in the
presence of M5-branes, which gives a dual description (for large number of
coincident branes) in terms of 11-dimensional supergravity on $AdS_7\times
S^4$.

\
\

The covariant field equations of an interacting (2,0) supermultiplet in
6-dimensions coupled to 11-dimensional supergravity background has been
found by Howe and Sezgin in \cite{er1}, which describe M5-brane
dynamics in M-theory. Since the conformal (2,0) theory in 6-dimensions is
also related to M5-brane dynamics, it is natural to expect a possible
connection to the equations of \cite{er1, er2}. Because the conformal
(2,0)
theory is not an ordinary field theory and has stringy excitations, a
connection with a field theory can only be established at ``large''
distances. Therefore, one should first find a way of breaking the 
conformal invariance in order to introduce a length scale in the theory.

\
\

The hint for how to break the conformal invariance comes from the
11-dimensional origin of the theory. It is well known that, in M-theory,
an M2-brane can end on an M5-brane \cite{str, townsend, becker, verlinde}.
This M2-brane
looks like a
(self-dual) string in 6-dimensional world-volume and preserves half of the
supersymmetries of the M5 brane theory. Thus, the states associated with
this self-dual string are stable, world-volume BPS states. The mass per
unit lenght of such a string would diverge, since it is the end of an
infinitely long M2-brane. However, it is possible to obtain finite energy
configurations simply by cutting the M2-brane at a finite length. In
11-dimensions this has the interpretation of placing another parallel
M5-brane for M2-brane to end. One can
now take a decoupling limit, where two M5-branes are still {\it separated}
from each other, but the mass of the streched membrane is fixed. The
resulting theory on the M5-brane is the (2,0) theory decoupled from the
bulk, however the conformal invariance is explicitly broken by the energy
scale of the streched membrane. This is similar to breaking of
the conformal
invariance of the ${\cal N}=4$, $U(N)$, Super Yang-Mills (SYM) theory by
separating a group of D3-branes from others (i.e., by Higgsing). In this
paper, we assume
that the (2,0) theory can be approximated by a field theory for distances
larger than the scale of the self-dual string tension, and the dynamics
is governed by the covariant equations obtained in \cite{er1}, where the
11-dimensional space is chosen to be flat.\footnote{To
be more precise as in the case of ${\cal N}=4$, $U(N)$ SYM, one should
first consider the equations of coincident M5-branes and break conformal
invariance by giving vacuum expectation values to some scalars. Since the
world-volume equations of the coincident M5-branes are not known, we
assume that at low energies, when the two parallel M5-branes are well
separated, one can consider the dynamics of each one independently.}

\
\

To support such a picture, the equations of the field theory should admit
a self-dual string background as a solitonic solution. This is indeed the
case; as shown in \cite{howe1} there is a supersymmetric self-dual string
solution which has an infinite mass per unit length. This is in agreement
with the interpretation of the string being the ending of a semi-infinite
membrane. As we will discuss in a moment, formally it is possible to
incorporate finite energy configurations simply by cutting the solution,
which can be thought of placing another parallel M5-brane for
M2-brane to end. Similar solutions describing fundamental-branes or
D-branes ending
on D-branes have been found in \cite{cal1, gib0}.

\
\

In the (2,0) theory (and in Dirac-Born-Infeld (DBI) theories in general),
fluctuations around a background configuration are characterized by the
so called {\it Boillat} metric \cite{gib, gibb2}. As we will discuss, the
Boillat metric for the self-dual string looks like a black-string metric
which has a singular horizon. Singularity of the horizon is associated
with the bad behaviour of the self-dual string solution at the origin. On
this background, one can consider a limit in which the
asymptotic region decouples from the "near horizon" region. As in the case
of
$AdS/CFT$ duality, this is possible thanks to a redshift factor related to
the underlying geometric structure of the theory. After taking the limit,
the
resulting background geometry can only be trusted for some region of the
radial parameter. This is very similar to non-conformal supergravity duals
of D-brane world-volume theories \cite{mal}.

\
\

A possible holographic dual for this system can be obtained by using an
effective string theory in 6-dimensions coupled to (2,0) tensor
multiplet. The world-sheet action for this string should have ${\cal
N}=(4,4)$ supersymmetry. Although the complete action is not known, one
can
easily determine the bosonic part (for a single string). In this action,
the coupling of the string world-sheet to the bulk (2,0) multiplet is
determined by the tension of the self-dual string. This implies that, in
the low energy limit that we will consider, the string action decouples
from
the bulk (2,0) theory. The decoupled theory can be identified with the
"near horizon" region of the self-dual string soliton, which would imply a
holographic duality.

\
\

The plan of the paper is as follows. In section 2, we discuss the (2,0)  
equations, the self-dual string solution and different possible 
descriptions of the system. In section 3, we define a low energy limit
which would imply a holographic duality between two theories. 
We conclude in section 4.

\section{The self-dual string solution}

\hs{5}Let us start with a brief description of the equations of
\cite{er1}, which are based on super-embedding of a 6-dimensional
world-volume superspace into 11-dimensional target superspace. The
equations are invariant under both bosonic and fermionic reprametrizations
of the world-volume superspace. The former can be regarded as the usual
diffeomerphism invariance and the latter is called the $\kappa$-symmetry.
The world-volume manifold has no specified metric structure, but it has an
induced geometry defined by the embedding map.

\
\

One can use super-reparametrization invariance of the world-volume to
choose the so called static gauge, where 6 of the 11 bosonic, and 16 of
the 32 fermionic embedding functions are identified with the bosonic and
fermionic world-volume coordinates respectively. This leaves 5 scalars, a
self-dual three-form and 16 fermionic fields as physical degrees of
freedom. After setting fermions to zero, the bosonic field equations are
given as \cite{er2}:
\bea\label{M5x}
G^{MN}\nabla_{M}\nabla_{N}X^{a'}=0,\\
G^{MN}\nabla_{M}H_{NPQ}=0, \label{M5h}
\eea
where $M,N=0,1,..5$, $X^{a'}$'s ($a'=1,..5$) represent transverse
fluctuations of the M5-brane and $H_{MNP}$ is the curl of the 2-form
potential. The covariant derivative $\nabla_{M}$ is constructed 
using $g_{MN}$, which is the
pull-back of the 11-dimensional metric into world-volume by the map
$X^{a'}$. 
The metric $G_{MN}$ is related to $g_{MN}$
by $E_{M}{}^{A}=e_{M}{}^{B}(m^{-1})_{B}{}^{A}$, where $E_{M}{}^{A}$ and
$e_{M}{}^{A}$ are vielbeins of $G_{MN}$ and $g_{MN}$ respectively,
and $m_{A}{}^{B}$ is defined as
\be
m_{A}{}^{B}=\delta_{A}{}^{B}-2h_{ACD}h^{BCD},
\ee
and 
$H_{MNP}=E_{M}{}^{A}E_{N}{}^{B}E_{P}{}^{C}m_{B}{}^{D}m_{C}{}^{E}h_{ADE}$.
Although, $H_{MNP}$ satisfies a non-linear self-duality condition, 
$h_{ABC}$ is linearly self-dual, and in
general it cannot be written as the curl of a two-form.

\
\

The equations (\ref{M5x}) and (\ref{M5h}) admit a supersymmetric self-dual
string solution \cite{howe1}, where the M5-brane is embedded into
11-dimensional flat Minkowski space. The solution is given by
\bea \label{eq1} 
X^{5'}&=&l_s\,h;\\ 
X^{i'}&=&\textrm{const.} \hs{5} \textrm{for} \hs{5} i'=1,2,3,4\\ 
H_{01m}&=&\frac{l_s}{4}\partial_{m}h,\\
H_{mnp}&=&\frac{l_s}{4}\e_{mnpq}\delta^{qr}\partial_{r}h \label{eq11}
\eea 
where
$\delta^{mn}\partial_{m}\partial_{n}h=0$ and thus 
\be \label{eq8}
h=1+\frac{q\,l_s^2}{r^{2}}. 
\ee 
In 6-dimensions, $(x^{0},x^{1})$ parametrize the string world-sheet, and
$y^{m}$ are the transverse coordinates. $X^{5'}$ is one of the coordinates
perpendicular to the M5-brane world-volume and together with $(x^{0},x^{1})$
it parametrizes the M2-brane ending on the M5-brane. The embedding
functions
$X^{a'}$, and the coordinates on the world-volume are chosen to have
dimension of length, and the components of the tensor $H_{MNP}$ are
dimensionless. The length scale $l_s$ and charge $q$ in
(\ref{eq8}) are related to the tension and the charge of the self-dual
string.

\
\

The mass per unit length of the solution can be calculated by first
performing a double dimensional reduction \cite{howe1}, which gives a
0-brane in 5-dimensions. Then, the energy is obtained by using the
DBI expression, and as shown in \cite{howe1, cal1}, it turns
out to be infinite. This is consistent with the interpretation that the
self-dual string is the ending of a semi-infinite M2-brane. From the
world-volume point of view, this infinity is associated with the $r\to 0$
behaviour of the solution, and it is similar to the self-energy divergence
of the Coulomb field in electromagnetism.

\
\

This shows that, the solution can only be trusted for $r>\e$ and should be
modified (or possibly the field equations can no longer be trusted) for
$r<\e$, where $\e$ is a small length scale. For the modified solution, the
energy should be finite. This has the interpretation of placing another
parallel M5-brane for M2-brane to end. Without knowing this modification, 
one could simply cut the space at $r=\e$. The energy would then have an
$\e$ dependence and thus $\e$ acts as a regulator in the
theory. (Following \cite{tow0}, \footnote{We thank D. Mateos for pointing 
out this reference to us.} one can sharpen these suggestions by
considering a test M5-brane in a supergravity M5-brane background. In this
case, the world-volume theory admits finite energy solutions
corresponding to M2-branes stretched between the test and the source
M5-branes \cite{tow0}.) 

\
\

It is possible to relate the length scale $l_s$ of the (2,0) theory 
and
the critical length $\e$ to the 11-dimensional planck length $l_p$ and the
separation of the M5-branes $L$. Since the solution should be modified for
$r<\e$, it is natural to assume that the M2-brane ends on the second
parallel M5-brane at $r=\e$. Noting
that the field $X^{5'}$ represents the transverse spike of the M2-brane,
one finds $L=X^{5'}(\e)$ (or at least $L$ and $X^{5'}(\e)$ have the same
order of magnitude). On the other hand, the energy (per unit length) of an
M2-brane with length $L$ is $T_2L$, where $T_{2}$ is the M2-brane tension.
This can be identified with the tension $T_s$ of the self-dual string, and
thus $1/l_s^2\sim T_s=T_2L$. Since $T_2\sim1/l_{p}^{3}$, $l_s$ can be
determined in terms of $l_p$ and $L$. For our purposes, we are interested
in a configuration where $\sqrt{q}l_s>>\e$. It is easy to see that this
corresponds to $L>>l_p$. In this case $l_s^2=l_p^3/L$ and
$\e^4=q^2\,\,l_p^9/L^5$.

\
\

At low energies, the dynamics of the self-dual string should be described by
an effective action involving the Goldstone modes of the solution, which
contains 4 bosonic and 4 fermionic zero modes. Although, we do not know
the
exact form of the action, it is possible to determine the bosonic part and 
its coupling to the bulk (2,0) fields.

\
\

In \cite{tow}, it was shown that there is a static membrane solution on
the M5-brane background of $D=11$ supergravity which can be interpreted as
the orthogonal intersection of an M2-brane with an M5-brane. In the same
paper, the dynamics of the membrane boundary on M5-brane was shown to
be governed by the Nambu-Goto action for a string moving in a
6-dimensional Minkowski space. To include non-trivial 6-dimensional bulk
excitations, one should consider a generalization of this action. The
fluctuations and causal structure around a non-trivial background of 
M5-brane fields are characterized by the {\it Boillat} metric
which is given by \cite{gibb2}
\be\label{boil}
C^{AB}=\frac{1}{K}\left[(1\,+\,\frac{4}{3}H^2) 
\,g^{AB}\,-\,4(H^2)^{AB}\right]
\ee
where
\be \label{K}
K=\sqrt{1+\frac{2}{3}H^2},
\ee
and all contractions are performed with the metric $g_{AB}$. Since the
zero
modes of a soliton can be viewed as fluctuations around that soliton,
it is natural to assume that the effective self-dual string ,which represents
the Goldstone modes, couples to the Boillat metric $C_{AB}$. \footnote{
$C_{AB}$ is defined to be the inverse of $C^{AB}$.} Therefore, one can 
write the following bosonic action
\be\label{str}
S_{string}=\frac{1}{l_s^2}\int d^2\sigma \,\,\sqrt{-\textrm{Det} 
\hat{C}}\,\,+\,\, \frac{q}{l_s^3} \int \,\,\hat{B},
\ee
where $B$ is the 2-form potential for the self-dual 3-form and $\hat{C}$ and 
$\hat{B}$ are pull-backs of the $C_{AB}$ and $B_{AB}$ to the 
world-sheet.

\
\

In this paper, we are mainly interested in how this string action couples
to the scalars of the (2,0) theory. This is important for understanding 
the emission, absorption or scattering of scalars by the effective string.
To see
the leading order couplings we first split 6-dimensional space $(z^{A})$
into $(x^\mu,y^m)$, where $\mu,\nu=0,1$ and $m,n=2,..,5$, and use
reparametrization invariance to identify the first two coordinates with
the world-volume coordinates of the string. We then expand the string
action (\ref{str}) to obtain
\be\label{str2}
S_{string}=\frac{1}{2l_s^2}\int d^2x \,\left[\,\,(\del_\mu X^{a'})^2 \,\,+\,\, 
(\del_\mu y^m)^2\right] + S'
\ee
where $y^m(x)$ are string world-sheet scalars and $S'$ starts with the 
quartic couplings between $\del_\mu X^{a'}$ and $\del_\mu y^{m}$ which 
have 
the following structure:
\be \label{s'}
S' \sim \frac{1}{l_s^2}\int d^2x\,\left[\,\del X\del X\del X\del X 
\,\,+\,\, \del X\del X\del y\del y \,\,+\,\, \del y\del y\del y\del y 
 + ....\right]
\ee
To include the M5-brane dynamics, this string action should be coupled to
the M5-brane action in the bulk. \footnote{As in the case of type IIB
supergravity, one can write an action and then impose the self-duality of
the 3-form field at the level of field equations.} The leading order
quadratic terms of such an action read
\be\label{bulk20}
S_{bulk}=\frac{1}{l_s^6}\int d^6z\,\, \left[\,\, (\del_A 
X^{a'})^2\,\,+\,\,(H_{ABC}H^{ABC})+....\right]
\ee
The total action governing the system is equal to
$S_{total}=S_{bulk}+S_{string}$.

\
\

Compared to the usual $p$-brane actions coupled to supergravity
scalars, $S_{total}$ has some interesting features. The most
striking difference is that there is an extra kinetic term for the bulk
scalars which appears on the world-sheet action (\ref{str2}). Also, the
structure of the interaction terms in (\ref{s'}) indicates that the bulk
scalars are {\it scattered} from the string rather than being absorbed or
emitted. This is mainly related to the fact that the bulk scalars are
themselves Goldstone modes of an M5-brane which represent transverse
fluctuations, and thus only the terms containing even number of $\del_\mu
X$ can appear in the interactions. All these are also consistent with the
fact that contrary to the $p$-branes of supergravity theories which are
{\it black} objects, the self-dual string is a soliton solution of a field
theory.

\
\

To see the zeroth order effect of the extra kinetic terms for the bulk
scalars, we ignore the interaction terms in $S_{total}$ and obtain the
following field equations\footnote{The field equations are
$[\del_A\del^A+\delta(y) \del_\mu\del^\mu]X^{a'}=0$. For smooth fields
each term should vanish separately which gives (\ref{bulk1}) and
(\ref{bulk2}).} for $X^{a'}$
\bea\label{bulk1}
\del_A\del^A X^{a'}&=&0,\\
\del_\mu\del^\mu X^{a'}|_{y=0}&=&0, \label{bulk2}
\eea
where we assume that the string world-sheet is located at $y^{m}=0$. For 
the localized modes that do not intersect the string,
equation (\ref{bulk2}) is 
satisfied identically and equation (\ref{bulk1}) gives usual 
massless modes propagating in the bulk. Also the other modes in the bulk
should obey (\ref{bulk2}), which implies \footnote{(\ref{bulk1}) gives 
usual massless modes $e^{i(-\omega t+ k_1x+k_my^m)}$, which are also 
eigenvectors of the operator $\del_\mu\del^\mu$ with non-zero eigenvalues. 
Therefore, (\ref{bulk2}) is equivalent to the Dirichlet boundary 
conditions $X^{a'}=0$ at $y=0$.}
\be\label{refl}
X^{a'}=e^{i(-\omega t+k_{1}x)}\sin(k_{m}y^{m}),
\ee
where $\omega^2=k_1^2+k_mk^m$. Thus, these modes scatter from the string
and form standing waves along transverse directions. This is the zeroth
order contribution to the scattering amplitude, and higher order
corrections can be calculated using the interaction terms in $S'$.

\
\

On the other hand, in finding the field equations, if one restricts the
field space so that they have non-zero support only on $y^m=0$ hyperplane,
then only $S_{string}$ gives non-zero contribution to the variation of
$S_{total}$, which implies
\be \label{free}
\del_\mu\del^\mu X^{a'} = 0. 
\ee
Since $X^{a'}$'s are assumed to have non-zero support only on $y^m=0$,
this
represents a massless mode purely {\it localized} on the string. It is
also possible to see the existence of this extra sector of modes from the 
path
integral approach. In such a functional integral, evaluating $S_{bulk}$
for the scalars with non-zero support only on the string world-sheet,
would give no contribution. However, $S_{string}$ gives non-zero
contrubution to the path integral, where the scalars also have the
standard kinetic term. This indicates existence of an extra massless 
sector in the theory.

\
\

Let us try to see the validity of the effective string picture by
comparing how scalars are scattered in the self-dual string solution
picture. Here we analyze the following scalar fluctuations around 
(\ref{eq1})-(\ref{eq11})
\be
X^{i'}=\,\textrm{const.}\,+\,\phi^{i'}, \hs{5} \textrm{for} 
\hs{5}i'=1,2,3,4.
\ee
and find that for all i' they obey the same equation
\be\label{lap}
\eta^{\mu\nu}\partial_{\mu}\partial_{\nu}\phi + f^{-1} 
\delta^{mn}\partial_{m}\partial_{n}\phi=0,
\ee
where 
\be \label{f}
f= 1\,+\,l_s^2\,\delta^{mn}\,\partial_{m}h\,\partial_{n}h=1+\frac{R^6}{r^{6}},
\ee
and $R^6=4q^{2}l_s^{6}$. Considering a spherically symmetric
s-wave, near the throat region the
constant term in (\ref{f}) can be ignored. In that case, one can define
a new radial coordinate by $\hat{r}\sim 1/r^2$ and equation
(\ref{lap}) becomes
\be 
(\del_x^2\,\,+\,\,\del_{\hat{r}}^2\,\,-\,\,\del_t^2)\phi=0.
\ee
This is nothing but the free-wave equation in the $t,x,\hat{r}$ plane. As
in the case of DBI theories, near the self-dual string, the solution is so
singular that one reaches another asymptotic region. From (\ref{eq1}), we
recognize that near $r=0$, $\hat{r}\sim X^{5'}(r)$, and thus these waves can
be considered to be propagating on the M2-brane. In the effective string
picture, this corresponds to the extra sector that we pointed out
following equation (\ref{free}). We see that the effective string picture
nicely captures the second asymptotic region.

\
\

Next, we consider the scattering of scalars from the self-dual string
solution. Since the potential term is so singular that near
$r=0$
there is another asymptotic region, we expect a perfect reflection at very
low energies. Considering a wave of the form $\phi(r,t)=e^{i\omega
t}\phi(r)$ we obtain a one-dimensional scattering problem given by

\be\label{onedim}
\left(\frac{d^2}{dz^2}\,\,+\,\,k(z)\right)\,\,\phi=0
\ee
where $z=(R\omega)^3/(r\omega)^2$ and 
\be 
k(z)=\frac{1}{4}[1+\frac{R^3\omega^3}{z^3}].
\ee
We would like to consider the problem in the limit $l_s\to0$ or
$R\omega\to 0$ which corresponds to a low energy scattering. To be able to
understand the structure of the potential in this limit, following
\cite{cal1}, we define a new coordinate suggested by the WKB
approximation
\be
x=\int_{\kappa}^{z}dz'\sqrt{k(z')},
\ee
where $\kappa=R\omega$. The coordinate $x$ extends from $-\infty$ to 
$+\infty$. Introducing a properly-normalized (in the sense of WKB) 
wave-function
\be
\phi=[k(z)]^{-1/4}\,\hat{\phi}
\ee
equation (\ref{onedim}) becomes
\be
\left(-\frac{d^2}{dx^2}+V(x)-1\right)\hat{\phi}=0, \hs{5} 
V(x)=\frac{a\kappa^{3}}{z^5k(z)^3}+\frac{b\kappa^6}{z^8k(z)^3},
\ee
where $a$ and $b$ are two positive numbers. This is a one-dimensional
quantum mechanical scattering problem with the potential $V(x)$. 
As $x\to\pm\infty$, $V(x)\to 0$ and $V(x)$ is finite at 
$x=0$. 

\
\

Let us try to see the form of the potential for low energy scattering when 
$l_s\to 0$, which implies $\kappa\to0$. By a scaling argument it is easy 
to see that $\int dx \kappa V(x)$ is independent of $\kappa$. Furthermore, 
$V(x)$ is everywhere positive and when $\kappa=0$, $V(x)$ is zero except 
at $x=0$. So, in the limit $\kappa\to 0$, $V(x)$ becomes a delta function. 
The scaling shows that
\be
V(x)\sim\frac{1}{\kappa}\delta(x),\hs{5}\kappa\to0.
\ee
The potential turns out to have the same limiting behaviour compared to
strings ending on D-branes and in this limit, one finds a {\it perfect}
reflection \cite{cal1}. Recall that in the effective string picture
in the same limit ($l_s\to0$), we had found perfectly reflecting modes
(\ref{refl}) from
the string. Thus, to the lowest order in perturbation theory,
the effective string picture nicely captures the scalar scattering
amplitudes on the string soliton.

\section{The low energy limit}

\hs{5}We now consider M-theory in the presence of two M5-branes and N
parallel M2-branes stretched between them. In 6-dimensions, this
corresponds to (2,0) theory in the presence of $N$ coincident self-dual
strings. At low energies, this system has two different
descriptions:

\
\

\hs{-4}i) from the world-volume theory point of view; one can consider the
self-dual string solution,

\
\

\hs{-4}ii) from the effective string picture; one can consider the string 
action in 6-dimensions describing the fluctuations of the 
$N$ coincident self-dual strings.

\
\

We would like to consider a limit where we let the tension of the
self-dual string $l_s\to 0$, and keep $L$ fixed, which corresponds to a
low energy limit. Since $l_p^3=Ll^2_s$, we also have $l_p\to 0$.
Therefore, 11-dimensional bulk excitations decouple from the system.
In the world-volume theory we would like to keep the energy of 
coincident strings finite. As pointed out in \cite{howe1}, the string
soliton might represent a D-string of (2,0) theory since the
``fundamental'' strings do not carry any $H$-charge 
\cite{verlinde, verlinde2}. Therefore, to keep the energy of the 
coincident strings finite, one should assume $r/l_s^2=\textrm{fixed}$ 
where $r$ is the separation between parallel self-dual strings.

\
\

Let us now apply the above limit to both descriptions of the system. The 
solution describing $N$ coincident self-dual strings is given by 
(\ref{eq1})-(\ref{eq11}) where the harmonic function $h$ is
\be
h=1+\frac{Nql_s^{2}}{r^{2}}.
\ee
In taking a low energy limit we need to be carefull how we define the
energy since the background solution induces a geometry. Of course far
from the string the induced geometry is flat and the
observer is located near that flat region. To determine how the
observer sees the string solution we look at the behaviour of small
fluctuations. In the previous section, we found that the scalar
perturbations obey equation (\ref{lap}). However, (\ref{lap}) is nothing
but the massless scalar equation on the background
\be\label{met}
ds^{2}=f^{-1/2}(-dt^{2}+dx^{2})\hs{2}+  
\hs{2}f^{1/2}(dr^{2}+r^{2}d\Omega_{3}^{2}),
\ee
which looks like a black-string in 6-dimensions.\footnote{For $N$
coincident strings, $R$ in (\ref{f}) is given by $R^6=4q^2N^2l_s^6$.} We
conclude that, although the theory does not have any dynamical gravity,
the induced geometry of the solution (\ref{eq1})-(\ref{eq11}) is such that
the observer far from the string sees it as a black-string background.

\
\

The same geometry can also be obtained by calculating the Boillat metric
(\ref{boil}) corresponding  to the self-dual string background. 
Actually this is not surprising, since the  Boillat metric
characterizes fluctuations around a non-trivial background. This
further indicates that a complete analysis of the scalar and 3-form 
fluctuations should reflect the geometry of (\ref{met}). 

\
\

It is important to remember that, the self-dual string soliton is defined 
for $r>\e$. Therefore, the above geometry can only be trusted for this
region. 
On the other hand, the metric (\ref{met}) has a horizon at $r=0$ which 
turns out to be a singular surface. As we will see shortly the horizon is 
conformal to $AdS_3\times S^3$. Although the solution is defined before 
reaching the horizon, there is still a large red-shift factor since $\e$ 
can in principle be chosen very small (but still non-zero) by moving the 
parallel M5-branes apart (thus letting $L$ to be very large). This
property of the  (2,0) theory makes it different from other ordinary field 
theories defined on flat spaces. 

\
\

We now consider the low energy limit on the self-dual string solution.
Remembering that we define the energies with respect to the observer
located at $r\to\infty$ and taking into account the red-shift factor in
the metric (\ref{met}), the low energy limit (which is equivalent to
taking $l_s\to 0$ with $r/l_s^{2}$ fixed\footnote{The situation is
very similar to the D-string of the Type IIB theory discussed in
\cite{mal}.}) 
gives the geometry
\be \label{limmet}
ds^{2}=\frac{r^{3}}{2Nq}(-dt^{2}+dx^{2})\hs{2}+\hs{2}\frac{2Nq}{r^{3}}dr^{2} 
\hs{2}+ \frac{2Nq}{r}\hs{2}d\Omega_{3}^{2}, 
\ee
where we set $l_s=1$. This is conformal to $AdS_{3}\times S^{3}$ and the
conformal factor is $1/r$. Now, recall that the geometry is defined only
for $r>\e$. On the other hand, the field equations break down at scales
smaller than $l_s$. Looking at the the curvature of (\ref{limmet}) one
finds that it is small compared to $l_s$ when $r<<Nq$. Therefore, for
large number of coincident strings, the low energy limit on the self dual
string can be characterized by the geometry of (\ref{limmet}), where the
radial coordinate is restricted to $\e<r<<Nq$. Note that the 
curvature gets smaller for larger $Nq$.

\
\

Let us now consider the same low energy limit in the effective string
picture. For $N$ coincident self-dual strings one should find the
non-abelian generalization of (\ref{str2}). This is a common problem for
DBI theories which has not been solved yet. On the other hand, from
(\ref{str2}) and (\ref{bulk20}), after properly normalizing the bulk
scalars, one finds that the coupling of the string world-sheet to the bulk
scalars and the extra kinetic terms of the bulk scalars on the world-sheet
vanishes as $l_s\to 0$. This indicates that in the low energy limit of the
non-abelian generalization of (\ref{str2}) one ends up with two different
theories; a non-trivial theory living on the string-world sheet and a
decoupled theory in the bulk. Similar to the well-known AdS/CFT duality,
we started with two different descriptions of the same system, and after
taking the limit we obtained two decoupled pieces in each description.
Identifying the two non-trivial pieces one can claim a {\it duality
between (2,0) theory on the background (\ref{limmet}) and a
two-dimensional field theory on the string}. The structure of
(\ref{limmet}) indicates that the string action is not conformal and for
some range of the renormalization group flow one can use the dual
geometry. Also, since the 6-dimensional space has 3 compact and 3
non-compact directions, (2,0) theory is effectively compactified to
3-dimensions. Thus, the suggested duality is a holographic duality in the
sense that a 3-dimensional theory (having diffeomorphism gauge invariance
and presumably tensionless strings) is dual to a 2-dimensional field
theory.

\
\

At this point, we need to be more explicit on what we mean by saying (2,0)
theory on the space defined by equation (\ref{limmet}). Since
(\ref{limmet}) does not have conformal symmetry, there is a length
scale in the theory. Thus, at low energies, we assume that the (2,0)
theory is described by the equations of \cite{er1}. Since these
equations are intrinsictly 11-dimensional, the most natural thing is
to start with a non-trivial background corresponding to M2-branes {\it
ending} on M5-branes, obtain the near horizon geometry (which corresponds
to a low energy description) and write down equations of \cite{er1} on
this geometry. Unfortunately, to the best of our knowledge, no 
{\it explicit}
supergravity solution describing M2-branes ending on an M5-brane exists 
in the
literature. On the other hand, the solution of M5-M2 brane {\it
intersection} is known. Choosing the harmonic functions of the branes
equal, the near horizon geometry of this intersection becomes
$AdS_{3}\times S^{3}\times R^{4}\times R$ (supported by non-trivial 4-form
fields). This should correspond to a conformal fixed point of the dual
theory. It would be interesting to find out branes {\it ending} on branes
type of solutions (which would presumably have a throat region given by
(\ref{limmet})) and give a precise meaning to the formulation of (2,0)
theory on the background of (\ref{limmet}).

\section{Conclusions}

\hs{4}In this paper, we considered the world-volume description of
M2-branes
ending on an M5-brane and defined a low energy limit which would imply a
holographic duality between two theories which do not contain gravity. The
duality could be established since the system has two different
descriptions. This is indeed an essential ingredient in all examples of
holography. Furthermore, in taking a low energy limit with respect to the
observer at infinity, one needs a mechanism for energies to be
red-shifted. Otherwise, in the low energy limit everything would become
trivial. In gravitational theories, existence of a horizon is responsible
for this red-shift. In our case, although we do not have a dynamical
gravity,
there is a hidden geometric structure. This defines a background geometry
for fluctuations around a field configuration and for the self-dual string
solution this geometry turns out to have a throat region. Therefore,
energies near the throat region are red-shifted and after the low energy
limit is taken one still ends up with non-trivial excitations.

\
\

The dualities proposed in this paper and in \cite{perr} are examples of
holography where dynamical gravity does not play any role. However, as
pointed out in the introduction, (2,0) theory is not an ordinary
field theory. Despite the absence of a dynamical gravity, in many
aspects it is
similar to a gravitational theory (since it is related to tensionless
strings). In the formulation of \cite{er1}, diffeomorphism invariance 
for both bosonic and fermionic fields plays a crucial role. Presumably, in
quantum theory, due to the diffeomorphism invariance acting as a gauge
symmetry, the degrees of freedom in the bulk may have a 
holographic description as in the case of gravitational theories.

\
\

As in \cite{perr}, one can also consider open membrane (OM)
theory in the same context. In \cite{perr}, OM metric was considered in
the large field limit, whereas in \cite{van, per3} the complete form
was found to be
\be\label{om}
(G_{OM})_{AB}=\frac{\left[(2K^2-1)-2K^2\sqrt{1-K^{-2}}\right]^{1/6}}{K}
\left(g_{AB}+4\,H^2_{AB}\right),
\ee
where K is given by (\ref{K}). Evaluating (\ref{om}) for the self-dual 
string solution (\ref{eq1})-(\ref{eq11}), one obtains the following metric
\be\label{ommet}
ds^{2}=f^{-2/3}(-dt^{2}+dx^{2})\hs{2}+  
\hs{2}f^{1/3}(dr^{2}+r^{2}d\Omega_{3}^{2}),
\ee
where $f$ is given by (\ref{f}). Although the solution
(\ref{eq1})-(\ref{eq11}) is singular at $r=0$ and thus valid for $r>\e$,
it turns out that (\ref{ommet}) is regular even at $r=0$. Indeed, $r=0$ is
a horizon and near horizon geometry is $AdS_3\times S^3$ without any
conformal factor. In \cite{perr}, a low energy limit is defined so that
the near horizon region is decoupled from the asymptotic region. Thus, one
has OM theory defined on $AdS_3\times S^3$ which would be dual to a
2-dimensional field theory. The dual $AdS_3$ geometry indicates that this
theory is defined at a conformal fixed point. The 2-dimensional field
theories considered in this paper and in \cite{perr} should be the same
(since they both describe collective motion of the same soliton). This
theory should have a renormalization group flow so that at an {\it
isolated} conformal fixed point it has a dual $AdS_3$ OM theory. When the
conformal invariance is broken, (\ref{limmet}) describes physics for some
interval in the renormalization group flow. It would be interesting to
verify such a picture. For that, one needs to find the complete ${\cal
N}=(4,4)$ supersymmetric action for $N$ coincident self-dual strings in
6-dimensions.

\
\

Finally, there is also a supersymmetric 3-brane soliton of the M5-brane
field equations \cite{howe3}. One may try to find the background geometry
corresponding to this 3-brane and define a low energy limit which would
imply a holographic duality. It turns out, the Boillat metric for the
3-brane does not have any horizon or throat region. It is a smooth metric
which interpolates between $R^6$ and $R^5\times S^1$. Thus, it is not
possible to define a decoupling limit in this case. The 3-brane in
6-dimensions has two transverse directions. Therefore, this object
requires a special treatment (like D7-brane of IIB theory) since it is not
a trivial task to define an asymptotic region.

\subsection*{Acknowledgements}
\hs{4}We are thankful to P. Sundell, D.S. Berman and T. Dereli for
valuable discussions. Basic ideas of this work took shape during
conversations with P. Sundell and D.S. Berman.


\begin{thebibliography}{25}

\bibitem{maldacena} J.M. Maldacena, {\sl The Large N Limit of
Superconformal
Field Theories and Supergravity}, Adv. Theor. Math. Phys. 2 (1998)
231, hep-th/9711200.

\bibitem{gubser} S.S. Gubser, I.R. Klebanov and A.M. Polyakov, {\sl Gauge
Theory Correlators from Non-Critical String Theory}, Phys. Lett.
B428 (1998) 105, hep-th/9802109.

\bibitem{witten} E. Witten, {\sl Anti-de Sitter Space and Holography},
Adv. Theor. Math. Phys. 2 (1998) 253, hep-th/9802150.

\bibitem{hooft}
G. 't Hooft, {\sl Dimensional Reduction in Quantum Gravity},
gr-qc/9310026.

\bibitem{susskind}
L. Susskind, {\sl The World as a Hologram}, J. Math. Phys. 36 (1995) 
6377, hep-th/9409089.

\bibitem{perr}
D.S. Berman and P. Sundell, {\sl $AdS_3$ OM Theory and the Self-Dual
String or 
Membranes Ending on the Five-Brane}, Phys. Lett. B529 (2002) 171, 
hep-th/0105288.

\bibitem{witm51}
E. Witten, {\sl Some Comments on String Dynamics}, hep-th 9507121.

\bibitem{witm52}
N. Seiberg and E. Witten, Comments on String Dynamics in 6-Dimensions, 
Nucl. Phys. B471 (1996) 121, hep-th 9603003.

\bibitem{seib1}
N. Seiberg, {\sl Notes on Theories with 16 Supercharges},  Nucl. 
Phys. Proc. Suppl. 67 (1998) 158, hep-th/9705117.

\bibitem{er1}
P.S. Howe and E. Sezgin, {\sl $D=11$, $p=5$}, Phys. Lett. B394 (1997) 62, 
hep-th/9611008.

\bibitem{er2}
P.S. Howe, E. Sezgin and P.C. West, {\sl Covariant Field Equations of the
M-Theory Five-Brane}, Phys. Lett. B399 (1997) 49, hep-th/9702008.             

\bibitem{str}
A. Strominger, {\sl Open p-Branes}, Phys. Lett. B383 (1996) 44,
hep-th/9512059. 

\bibitem{townsend}
P.K. Townsend, {\sl D-Branes from M-Branes}, Phys. Lett. B373 (1996) 68,
hep-th/9512062.

\bibitem{becker}
K. Becker and M. Becker, {\sl Boundaries in M-Theory}, Nucl.Phys. B472
(1996) 221, hep-th/9602071.

\bibitem{verlinde}
R. Dijkgraaf, E. Verlinde and H. Verlinde, {\sl BPS Spectrum of the
Five-Brane and Black Hole Entropy}, Nucl.Phys. B486 (1997) 77,
hep-th/9603126.

\bibitem{howe1}
P.S. Howe, N.D. Lambert and P.C. West, {\sl The Self-Dual String Soliton}, 
Nucl. Phys. B515 (1998) 203, hep-th/9709014.

\bibitem{cal1}
C.G. Callan and J.M. Maldacena, {\sl Brane Dynamics from the Born-Infeld 
Action}, Nucl. Phys. B513 (1998) 198, hep-th/9708147.

\bibitem{gib0}
G.W. Gibbons, {\sl Born-Infeld Particles and Dirichlet p-Branes},  Nucl.
Phys. 
B514 (1998) 603, hep-th/9709027.

\bibitem{gib}
G.W. Gibbons, {\sl Aspects of Born-Infeld Theory and String/M-Theory},  
hep-th/0106059.

\bibitem{gibb2}
G.W. Gibbons and P.C. West, {\sl The Metric and Strong Coupling Limit of
the 
M5-Brane}, J. Math. Phys. 42 (2001) 3188, hep-th/0011149.

\bibitem{mal}
N. Itzhaki, J.M. Maldacena, J. Sonnenschein and S. Yankielowicz, 
{\sl Supergravity and The Large N Limit of Theories with Sixteen
Supercharges}. 
Phys. Rev. D58 (1998) 046004, hep-th/9802042.

\bibitem{tow0}
J.P. Gauntlett, C. Koehl, D. Mateos, P.K. Townsend and M. Zamaklar,
{\sl Finite Energy Dirac-Born-Infeld Monopoles and String Junctions}, 
Phys. Rev. D60 (1999) 045004, hep-th/9903156.

\bibitem{tow}
J. Gomis, D. Mateos, J. Simon and P.K. Townsend, {\sl Brane Intersection  
Dynamics from Branes Ending in Brane Backgrounds}, Phys. Lett. B430 (1998) 
231, hep-th/9803040.

\bibitem{verlinde2}
R. Dijkgraaf, E. Verlinde and H. Verlinde, {\sl BPS Quantization of 
the Five-Brane}, Nucl. Phys. B486 (1997) 89, hep-th/9604055. 

\bibitem{van}
J.P. van der Schaar, {\sl The Reduced Open Membrane Metric}, JHEP 0108
(2001) 048, hep-th/0106046.

\bibitem{per3}
D.S. Berman, M. Cederwall, U. Gran, H. Larsson, M. Nielsen,
B.E.W. Nilsson and P. Sundell, {\sl Deformation Independent Open Brane
Metrics and Generalized Theta Parameters}, JHEP 0202 (2002) 012, 
hep-th/0109107.

\bibitem{howe3}
P.S. Howe, N.D. Lambert and P.C. West, {\sl The Threebrane Soliton of the
M-Fivebrane}, Phys. Lett. B419 (1998) 79, hep-th/9710033.


\end{thebibliography}
\end{document}